\documentstyle[12pt,fleqn]{article}
\def\la{\langle}
\def\ra{\rangle}
\def\beeq{\begin{equation}}
\def\eneq{\end{equation}}
\def\beeqa{\begin{eqnarray}}
\def\eneqa{\end{eqnarray}}

\addtocounter{section}{-1}
\begin{document}

\begin{center}

\vspace{2cm}

{\large {\bf {Optical excitations in electroluminescent polymers:\\
poly({\mbox{\boldmath $para$}}-phenylenevinylene) family\\
} } }

\vspace{1cm}
(Running head: {\sl Optical excitations in PPV family})

\vspace{1cm}

{\rm Kikuo Harigaya\footnote[1]{E-mail address: 
\verb+harigaya@etl.go.jp+; URL: 
\verb+http://www.etl.go.jp/People/harigaya/+}}

\vspace{1cm}

{\sl Physical Science Division,
Electrotechnical Laboratory,\\ 
Umezono 1-1-4, Tsukuba, Ibaraki 305, Japan}

\vspace{1cm}

(Received~~~~~~~~~~~~~~~~~~~~~~~~~~~~~~~~~~~)
\end{center}

\vspace{1cm}

\noindent
{\bf Abstract}\\
Component of photoexcited states with large spatial extent 
is investigated for optical absorption spectra of the 
electroluminescent conjugated polymers by using the intermediate 
exciton theory.  We calculate the ratio of oscillator strengths 
due to long-range excitons with respect to sum of all the oscillator 
strengths of the absorption as a function of the monomer number.  
The oscillator strengths of the long-range excitons in 
poly({\sl para}-phenylene) are smaller than those in 
poly({\sl para}-\-phen\-yl\-ene\-vinylene), and those of 
poly({\sl para}-\-phen\-yl\-ene\-di\-vinylene) are
larger than those in poly({\sl para}-\-phen\-yl\-ene\-vinylene).  
Such relative variations are explained by the
differences of the number of vinylene units.
The oscillator strengths of long-range excitons in 
poly(di-{\sl para}-\-phen\-yl\-ene\-vinylene)
are much larger than those of the above three polymers,
due to the increase of number of phenyl rings.
We also find that the energy position of the almost localized 
exciton is neary the same in the four polymers.

\mbox{}


\pagebreak

\section{Introduction}

The observation of remarkable electroluminescent properties
of the polymer poly({\sl para}-\-phen\-yl\-ene\-vinylene) (PPV) [1] 
has attracted physical and chemical research activities.  An 
important recent development is the observation of stimulated 
emission and lasing from a PPV layer incorporated within an 
optical cavity [2].  Lasing has been also proposed in order to 
explain the band-narrowing observed on high intensity 
photoexcitation of a composite film of a PPV derivative and 
nanoparticles of TiO$_2$ [3].  Therefore, understanding of 
structures of photoexcited states in PPV and related polymers 
containing phenyl rings is one of interesting research topics.

In the PPV polymer, the onset of the photocurrents locates at the 
excitation energy between 3.0 eV and 4.0 eV [4,5,6], and this 
energy is significantly larger than both of the optical absorption 
edge at about 2.0 eV and the lowest peak energy at 2.4 eV.  
In the previous study [7,8], we have characterized the extent of 
photoexcited states of the PPV by using the intermediate exciton
theory.  When the distance between an electron and a hole is 
shorter than the spatial extent of the monomer, we have called 
the exciton as a ``short-range" exciton.  When the exciton width 
is larger than the extent of the monomer, we have called the 
exciton as a ``long-range" exciton.  We have characterized each 
photoexcited state as ``short-range" or ``long-range".  We have 
shown that a long-range exciton feature starts at the energy in 
the higher energy side of the lowest feature of the optical 
absorption of PPV.  The energy position is nearly the same as 
that of the semiconducting energy gap which has been interpreted 
as the onset of the large photocurrents observed in experiments.
Therefore, the presence of photoexcited states with large spatial
extent is essential in mechanisms of remarkable photocurrents
observed in PPV.

The purpose of this paper is to look at contributions from 
long-range excitons, which make large effects on photoconduction
properties in PPV-related polymers: poly({\sl para}-phenylene) 
(PPP), poly({\sl para}-phenylenedivinylene) (PPD), and 
poly(di-{\sl para}-phenylenevinylene) (PDV).  The structures
of these polymers, namely, PPV family, are illustrated in Fig. 1.
These polymers are composed of phenyl rings and vinylene CH-dimers.
The names of the two polymers and their abbreviations, PPD and PDV,
are pseudonyms (tentative names) in this paper, because current 
common names of these polymers have not been known at present.  
The long-range excitons are characterized as we have done for 
PPV in the previous study [7,8].

In the next section, the theoretical formalism is explained,
and the characterization method of long-range excitons is given.
The calculated results are reported in \S 3, and the paper is 
concluded with a summary in \S 4.

\section{Model}

We consider the following model with electron-phonon and 
electron-electron interactions.  
\beeqa
H &=& H_{\rm pol} + H_{\rm int}, \\
H_{\rm pol} &=& - \sum_{\la i,j \ra,\sigma} ( t - \alpha y_{i,j} )
( c_{i,\sigma}^\dagger c_{j,\sigma} + {\rm h.c.} )
+ \frac{K}{2} \sum_{\la i,j \ra} y_{i,j}^2, \\
H_{\rm int} &=& U \sum_{i} 
(c_{i,\uparrow}^\dagger c_{i,\uparrow} - \frac{n_{\rm el}}{2})
(c_{i,\downarrow}^\dagger c_{i,\downarrow} 
- \frac{n_{\rm el}}{2}) \nonumber \\
&+& \sum_{i,j} W(r_{i,j}) 
(\sum_\sigma c_{i,\sigma}^\dagger c_{i,\sigma} - n_{\rm el})
(\sum_\tau c_{j,\tau}^\dagger c_{j,\tau} - n_{\rm el}).
\eneqa
In eq. (1), the first term $H_{\rm pol}$ is the tight 
binding model along the polymer backbone with electron-phonon 
interactions which couple 
electrons with modulation modes of the bond lengths, and 
the second term $H_{\rm int}$ is the Coulomb interaction 
potentials among electrons.  In eq. (2), $t$ is the 
hopping integral between the nearest neighbor carbon atoms 
in the ideal system without bond alternations; $\alpha$ 
is the electron-phonon coupling constant that modulates 
the hopping integral linearly with respect to the bond 
variable $y_{i,j}$ which measures the magnitude of the 
bond alternation of the bond $\la i,j \ra$; $y_{i,j} > 0$ 
for longer bonds and $y_{i,j} < 0$ for shorter bonds (the 
average of $y_{i,j}$ is taken to be zero); $K$ is the 
harmonic spring constant for $y_{i,j}$; and the sum is 
taken over the pairs of neighboring atoms.  Equation (3) is 
the Coulomb interactions among electrons.  Here, $n_{\rm el}$ 
is the average number of electrons per site; $r_{i,j}$ is 
the distance between the $i$th and $j$th sites; and 
\beeq
W(r) = \frac{1}{\sqrt{(1/U)^2 + (r/a V)^2}}
\eneq
is the parametrized Ohno potential.  The quantity $W(0) = U$ 
is the strength of the onsite interaction; $V$ means the 
strength of the long-range part ($W(r) \sim aV/r$ in the limit 
$r \gg a$); and $a = 1.4$\AA~ is the mean bond length.  The 
parameter values used in this paper are $\alpha = 2.59t$/\AA, 
$K=26.6t$/\AA$^2$, $U=2.5t$, and $V=1.3t$.  They have been 
determined by comparison with experiments of PPV, and have 
been used in [7,8].  Most of the quantities in the energy units 
are shown by the unit of $t$ in this paper.

Excitation wavefunctions of the electron-hole pair are 
calculated by the Hartree-Fock approximation followed 
by the single excitation configuration interaction method.  
This method, which is appropriate
for the cases of moderate Coulomb interactions -- strengths 
between negligible and strong Coulomb interactions -- is 
known as the intermediate exciton theory in the literatures [7,8].
We write the singlet electron-hole excitations as
\beeq
|\mu, \lambda \rangle = \frac{1}{\sqrt{2}}
(c_{\mu, \uparrow}^\dagger c_{\lambda, \uparrow} 
+ c_{\mu, \downarrow}^\dagger c_{\lambda, \downarrow} )
| g \rangle,
\eneq
where $\mu$ and $\lambda$ mean unoccupied and occupied states, 
respectively, and $| g \rangle$ is the Hartree-Fock ground state.  
The general expression of the $\kappa$th optical excitation is:
\beeq
| \kappa \rangle = \sum_{(\mu,\lambda)} D_{\kappa,(\mu,\lambda)}
| \mu, \lambda \rangle.
\eneq
After inserting the relation with the site representation 
$c_{\mu,\sigma} = \sum_i \alpha_{\mu,i} c_{i,\sigma}$, we obtain
\beeq
| \kappa \rangle = \frac{1}{\sqrt{2}} \sum_{(i,j)} 
B_{\kappa,(i,j)} (c_{i,\uparrow}^\dagger c_{j,\uparrow} 
+ c_{i,\downarrow}^\dagger c_{j,\downarrow} ) | g \rangle,
\eneq
where 
\beeq
B_{\kappa,(i,j)} = \sum_{(\mu,\lambda)} D_{\kappa,(\mu,\lambda)}
\alpha_{\mu,i}^* \alpha_{\lambda,j}.
\eneq
Thus, $|B_{\kappa,(i,j)}|^2$ is the probability that an electron 
locates at the $i$th site and a hole is at the $j$th site.

We shall define the following quantity:
\beeq
P_\kappa = \sum_{i \in M} \sum_{j \in M} |B_{\kappa,(i,j)}|^2,
\eneq
where $M$ is a set of sites within a single monomer, in other
words, a set of carbon sites included in the brackets of each
polymer shown in Fig. 1.  When $P_\kappa > 1/N_{\rm m}$ 
($N_{\rm m}$ is the number of monomers used in the calculation
of periodic polymer chains), the electron and hole favor to 
have large amplitudes in the same single monomer.  Then, this 
excited state is identified as a short-range exciton.  On the 
other hand, when $P_\kappa < 1/N_{\rm m}$, the excited state is 
characterized as a long-range exciton.  This characterization 
method is performed for all the photoexcited states 
$| \kappa \rangle$, and a long-range component in the optical 
absorption spectrum is extracted from the total absorption.  
In the next section, numerical results are to be reported and 
discussion will be given.

\section{Long-range excitons in PPV family}
\subsection{PPP and PPV}

We discuss properties of photoexcited states in PPP [Fig. 1 (a)] 
and PPV [Fig. 1 (b)] in this subsection.  Numerical results of 
PPV have been shown in refs. [7,8] already, but they are shown again 
comparing with the results of PPP.  Figures 2 and 3 show optical 
absorption spectra and long-range component of the oscillator 
strengths for PPP, and Figs. 4 and 5 show results of PPV.

Figures 2 (a) and (b) show anisotropies of the optical absorption
of PPP.  The electric field of light is parallel and perpendicular
to the polymer axis in Figs. 2 (a) and (b), respectively.  The 
thin lines show the contributions from long-range excitons.  In 
Fig. 2 (a), there are two features centered around about $1.4t$ 
and $2.5t$.  The former comes from optical excitations between 
the highest occupied molecular orbital (HOMO) and the lowest 
unoccupied molecular orbital (LUMO).  The HOMO and LUMO have 
certain magnitudes of dispersions near the Brillouin zone center, 
and therefore the excitons near the energy $1.4t$ have an 
appreciable long-range component.  In fact, we observe a peak 
due to long-range excitons around the energy $1.8t$.  On the 
other hand, the feature around $2.5t$ is due to transitions between 
spatially localized band states [9].  So, this feature does not 
have an appreciable long-range component in the higher energy 
side of the peak.  Next, we discuss the case with perpendicular 
electric field shown in Fig. 2 (b).  There is a feature at
$2.2t$, whose long-range exciton component exists at the higher
energy side.  In higher energies than this feature, several
transitions are mixed, and the long-range component is dominant
in energies larger than $3.0t$.  Figure 2 (c) shows the 
absorption spectra where the direction of the electric field 
of light is orientationally averaged.  We find that two features
around $1.4t$ and $2.5t$ in the parallel electric field case 
give the central two peaks in the spectra.  The features in
the perpendicular field case overlap with the lower and 
higher energy sides of the dominant feature around $2.5t$.

Figure 3 summarizes the long-range component of the oscillator
strengths of PPP.  It is shown as a function of the PPP monomer 
number $N_{\rm m}$.  The squares are for the total absorption.  
The circles and triangles indicate the data for the cases with 
the electric field parallel and perpendicular to the polymer 
axis, respectively.  Generally, the long-range component of
the perpendicular electric field case is larger than that of the 
parallel field case.  It seems that the long-component of the
orientationally averaged spectra saturates at approximately 
7 \% near $N_{\rm m} = 20$.  Note that the small fluctuation 
of the plots with respect to $N_{\rm m}$ is due to the numerical
calculation errors, and is not an intrinsic behavior.  Such 
numerical fluctuations have also been seen in our previous
work of PPV [7,8].  Readers are suggest to look at the overall 
variations of the long-range component with respect to $N_{\rm m}$.

Next, we compare the results of PPP with those of PPV.  
Numerical absorption spectra are shown in Fig. 4, and their 
long-range component is shown in Fig. 5.  The results have
been already reported in [7,8], but they are shown again 
in order to compare with results of the other polymers.
Comparing Fig. 2 with Fig. 4, we observe that peak structures
of PPP and PPV are similar at the first glance.  However, 
there is a quantitative difference: the oscillator strengths
of the lowest excitation in Figs. 4 (a) and (c) are larger 
in PPV than those of PPP.  The long-range component of the 
lowest exciton is larger in PPV, too.  This fact reflects 
in the larger saturated value of the total long-range component
of PPV shown in Fig. 5 than that of PPP shown in Fig. 3.

As noted in ref. [9], the lowest optical excitations of PPP 
and PPV have characters like those of excitons in a simple 
prototype polymer: {\sl trans}-polyacetylene.  The band 
structures of polyacetylene are well described by the 
Su-Schrieffer-Heeger (SSH) model [10].  The quasiparticle band 
structures have finite dispersions at the top (bottom) of the 
valence band (conduction band).  Therefore, the lowest 
exciton owing to these dispersive bands has apparent 
long-range component at the higher energy side after taking
into account of Coulomb interactions among electrons.
The lowest excitons of PPP and PPV have just the similar
characters, and thus Figs. 2 (a) and 4 (a) show the dominant
contributions of long-range excitons with respect to the 
lowest band-to-band optical excitations.

The difference in the polymer structures of PPP and PPV 
in Figs. 1 (a) and (b) is that a vinylene bond is added
between phenyl rings of PPP.  This is related with the 
property that electronic structures near the energy gap 
of PPV have more dispersive characters than those of PPP, 
and thus the optical excitations owing to these bands will 
more resemble with those of {\sl trans}-polyacetylene 
described by the SSH model.  Therefore, the relative 
oscillator strengths of the lowest excitons of PPV 
[Fig. 4 (a)] are larger than those of PPP [Fig. 2 (a)]. 
Also, total long-range component (about 8 \%) of PPV 
(Fig. 5) is larger than that (about 7 \%) of PPP (Fig. 3).  
We could say that optical excitations of PPV will have 
photoconductive properties which are stronger than those of PPP.

\subsection{PPD}

In PPD shown in Fig. 1 (c), there are two vinylene bonds 
between the neighboring phenyl rings.  In this subsection,
we look at the optical properties resulting from electronic
structures and excitonic effects of PPD.

Figure 6 shows the optical absorption spectra, when the 
electric field of light is parallel to the polymer axis 
[Fig. (a)], perpendicular to the polymer [Fig. (b)], and 
is averaged over orientations [Fig. (c)].  Long-range 
exciton contributions are also shown in these three figures.
When the electric field is along with the polymer axis
[Fig. 6 (a)], the oscillator strengths of the lowest
exciton around the energy $1.2t$ are dominant.  This feature
has its long-range component at the higher energy side.
There is a small feature around the energy $2.5t$.
The oscillator strengths of this feature becomes relatively
smaller than those of PPV shown in Fig. 4 (a).  The lowest
exciton feature becomes larger than that of Fig. 4 (a).
This comes from the larger number of vinylene units in PPD
than that of PPV.  When the electric field is perpendicular
to the polymer [Fig. 6 (b)], the overall spectral shape is
similar to that in Fig. 4 (b).   Figure 6 (c) shows the 
orientationally averaged spectra.  The lowest exciton
feature becomes more dominant from PPP [Fig. 2 (c)], PPV [Fig. 4 (c)],
to PPD.  The three peak structure from $\sim 2t$ to $\sim 3t$
is commonly seen in these three polymers, but their 
oscillator strengths become relatively weaker.

Figure 7 shows the long-range component of PPD as a function
of the monomer number $N_{\rm m}$.  The maximum number of 
$N_{\rm m}$ is 12.  The saturation behavior is seen at smaller
$N_{\rm m}$ than that of PPV, due to the larger dimension
of the monomer unit.  The long-range component of the perpendicular
field case is larger than that of the parallel field case.
The long-range component of the orientationally averaged case
saturates at $N_{\rm m} \sim 10$ with the value of approximately
11 \%.  This polymer will be more photoconductive than PPV.

\subsection{PDV}

In previous subsections, we have considered the series:
PPP, PPV, and PPD, where number of vinylene bonds increases
in each monomer unit.  There would be another series:
PPP, PPV, and PDV [shown in Fig. 1 (d)], where number of
phenyl rings increases in the monomer units.  We consider
optical properties of PDV in this subsection.

Figure 8 (a) shows the optical absorption spectra of the 
parallel electric field case.  There are two main features:
the lowest exciton peak around $1.0t$ and the almost localized 
exciton feature around $2.5t$.  They are commonly seen in 
polymers considered in this paper.  The broad structure from 
$\sim 3t$ to $\sim 4t$ is due to mixing of high energy
transitions among molecular orbitals with large energies in the 
presence of more phenyl rings.  Figure 8 (b) is the spectra
where the electric field is perpendicular to the polymer.
In contrast to the parallel field case, the spectral shape
is rather similar to that of the other polymers.  The long-range
exciton contributions are remarkable in this case.  
Figure 8 (c) shows the averaged spectra.  The overall shape
is like that of PPV shown in Fig. 4 (c).  There is a distinct
lowest feature at around $1.0t$, and three small peaks are
present in higher energies.  However, the long-range 
component is larger than that of PPV.  This polymer will
show the largest photoconductive properties among polymers
considered in this paper.

Finally, Fig. 9 shows the long-range component as a function
of the number of monomer units.   The long-range component
of the perpendicular field case is larger than that of the 
parallel field case, as clearly seen in Fig. 8 (b).  The 
saturated value of the orientationally averaged spectra is 
approximately 17 \%.  This is due to the complex molecular 
orbital structures of PDV.  Table I summarizes the long-range 
component of the four polymers before closing this section.

\section{Summary}

We have considered optical excited states of the electroluminescent
polymers which are related with the most famous polymer PPV.
We have payed attention to the long-range component of the 
optical excitations using the intermediate exciton theory.
The following is the main conclusions of this paper:\\
(1) As we go from PPP, PPV, to PPD, the oscillator strengths 
of the long-range excitons become larger.  Such the relative 
variation is explained by the difference of number of the 
vinylene bonds in each monomer unit.\\
(2) The oscillator strengths of long-range excitons in PDV
are much larger than those of the above three polymers, due 
to the increase of number of phenyl rings.\\
(3)  The energy position of the almost localized exciton is 
neary the same ($\sim 2.5t$) in the four polymers considered 
in this paper.  The excitation energy of the almost localized 
exciton is just $2.0t$ in a free electron model [9] for all 
the polymers.  The enhancement about $0.5t$ originates from 
the Coulomb interaction part of our model.

\mbox{}

\begin{flushleft}
{\bf Acknowledgements}
\end{flushleft}

Useful discussion with Y. Shimoi, S. Abe, S. Kobayashi, 
K. Murata, and S. Kuroda is acknowledged.  Helpful 
communications with S. Mazumdar, M. Chandross, W. Barford, 
D. D. C. Bradley, R. H. Friend, E. M. Conwell, and 
Z. V. Vardeny are much acknowledged.

\pagebreak
\begin{flushleft}
{\bf References}
\end{flushleft}

\noindent
$[1]$ J. H. Burroughes, D. D. C. Bradley, A. R. Brown, R. N. Marks,
K. Mackay, R. H. Friend, P. L. Burns, and A. B. Holmes, 
Nature {\bf 347}, 539 (1990).\\ 
$[2]$ N. Tessler, G. J. Denton, and R. H. Friend, Nature {\bf 382},
695 (1996).\\
$[3]$ F. Hide, B. J. Schwarz, M. A. Diaz-Garcia, and A. J. Heeger,
Chem. Phys. Lett. {\bf 256}, 424 (1996).\\
$[4]$ K. Pichler, D. A. Halliday, D. D. C. Bradley, P. L. Burn,
R. H. Friend, and A. B. Holmes, J. Phys.: Condens. Matter {\bf 5}, 
7155 (1993).\\
$[5]$ D. A. Halliday, P. L. Burn, R. H. Friend, D. D. C. Bradley,
A. B. Holmes, and A. Kraft,  Synth. Met. {\bf 55-57}, 954 (1993).\\
$[6]$ K. Murata, S. Kuroda, Y. Shimoi, S. Abe, T. Noguchi,
and T. Ohnishi, J. Phys. Soc. Jpn. {\bf 65}, 3743 (1996).\\
$[7]$ K. Harigaya, Y. Shmoi, and S. Abe, in {\sl Proceedings
of the 2nd Asia Symposium on Condensed Matter Phytophysics}
(Nara, 1996), p. 25.\\
$[8]$ K. Harigaya, J. Phys. Soc. Jpn. {\bf 66}, to be published (1997).\\
$[9]$ Z. G. Soos, S. Etemad, D. S. Galv\~{a}o, and S. Ramasessha,
Chem. Phys. Lett. {\bf 194}, 341 (1992).\\
$[10]$ W. P. Su, J. R. Schrieffer, and A. J. Heeger, Phys. Rev. B
{\bf 22}, 2099 (1980).\\

\pagebreak

\noindent
TABLE I.  The component of long-range excitons 
in four kinds of polymers for which we calculated.

\mbox{}

\begin{tabular}{cc} \hline \hline
polymer & long-range component \\ \hline
PPP     & 7 \% \\
PPV     & 8 \% \\
PPD     & 11 \% \\
PDV     & 17 \% \\ \hline \hline
\end{tabular}

\pagebreak

\begin{flushleft}
{\bf Figure Captions}
\end{flushleft}

\mbox{}

\noindent
Fig. 1.  Lattice structures of the electroluminescent
polymers studied in this paper.  The abbreviations, 
PPD and PDV, are pseudonyms.

\mbox{}

\noindent
Fig. 2. Optical absorption spectra of the PPP.  The polymer 
axis is in the $x$-$y$ plane.   The electric field of light 
is parallel to the chain and in the direction of the $x$-axis 
in (a), and it is perpendicular to the axis and is along 
with the $z$-axis in (b).  The orientationally averaged spectra 
are shown in (c).   The number of the PPP units is $N_{\rm m}= 19$.  
The bold line is for the total absorption.  The thin line 
indicates the absorption of the long-range component.  
The Lorentzian broadening $\gamma = 0.15 t$ is used.

\mbox{}

\noindent
Fig. 3.  Long-range component of the optical absorption
spectra as a function of the PPP unit number $N_{\rm m}$.
The squares are for the total absorption.  The circles 
and triangles indicate the data for the cases with the 
electric field parallel and perpendicular to the polymer 
axis, respectively.

\mbox{}

\noindent
Fig. 4. Optical absorption spectra of the PPV.  The polymer 
axis is in the $x$-$y$ plane.   The electric field of light 
is parallel to the chain and in the direction of the $x$-axis 
in (a), and it is perpendicular to the axis and is along 
with the $z$-axis in (b).  The orientationally averaged spectra 
are shown in (c).   The number of the PPV units is $N_{\rm m}= 20$.  
The bold line is for the total absorption.  The thin line 
indicates the absorption of the long-range component.  
The Lorentzian broadening $\gamma = 0.15 t$ is used.

\mbox{}

\noindent
Fig. 5.  Long-range component of the optical absorption
spectra as a function of the PPV unit number $N_{\rm m}$.
The squares are for the total absorption.  The circles 
and triangles indicate the data for the cases with the 
electric field parallel and perpendicular to the polymer 
axis, respectively.

\mbox{}

\noindent
Fig. 6. Optical absorption spectra of the PPD.  The polymer 
axis is in the $x$-$y$ plane.   The electric field of light 
is parallel to the chain and in the direction of the $x$-axis 
in (a), and it is perpendicular to the axis and is along 
with the $z$-axis in (b).  The orientationally averaged spectra 
are shown in (c).   The number of the PPD units is $N_{\rm m}= 12$.  
The bold line is for the total absorption.  The thin line 
indicates the absorption of the long-range component.  
The Lorentzian broadening $\gamma = 0.15 t$ is used.

\mbox{}

\noindent
Fig. 7.  Long-range component of the optical absorption
spectra as a function of the PPD unit number $N_{\rm m}$.
The squares are for the total absorption.  The circles 
and triangles indicate the data for the cases with the 
electric field parallel and perpendicular to the polymer 
axis, respectively.

\mbox{}

\noindent
Fig. 8. Optical absorption spectra of the PDV.  The polymer 
axis is in the $x$-$y$ plane.   The electric field of light 
is parallel to the chain and in the direction of the $x$-axis 
in (a), and it is perpendicular to the axis and is along 
with the $z$-axis in (b).  The orientationally averaged spectra 
are shown in (c).   The number of the PDV units is $N_{\rm m}= 9$.  
The bold line is for the total absorption.  The thin line 
indicates the absorption of the long-range component.  
The Lorentzian broadening $\gamma = 0.15 t$ is used.

\mbox{}

\noindent
Fig. 9.  Long-range component of the optical absorption
spectra as a function of the PDV unit number $N_{\rm m}$.
The squares are for the total absorption.  The circles 
and triangles indicate the data for the cases with the 
electric field parallel and perpendicular to the polymer 
axis, respectively.

\end{document}